# Direct Measurement of Exciton Valley Coherence in Monolayer WSe$_2$

Kai Hao[1,*], Galan Moody[1,†,*], Fengcheng Wu[1], Chandriker Kavir Dass[1], Lixiang Xu[1], Chang-Hsiao Chen[2], Ming-Yang Li[3], Lain-Jong Li[3], Allan H. MacDonald[1] and Xiaoqin Li[1]

[1] Department of Physics and Center for Complex Quantum Systems, University of Texas at Austin, Austin, TX 78712, USA.

[2] Department of Automatic Control Engineering, Feng Chia University, Taichung 40724, Taiwan.

[3] Physical Science and Engineering Division, King Abdullah University of Science & Technology (KAUST), Thuwal 23955, Saudi Arabia.

[†]Present address: National Institute of Standards & Technology, Boulder, CO 80305, USA.

[*]These authors contributed equally to this work.

**In crystals, energy band extrema in momentum space can be identified by their valley index. The internal quantum degree of freedom associated with valley pseudospin indices can act as a useful information carrier analogous to electronic charge or spin. Interest in valleytronics has been revived in recent years following the discovery of atomically thin materials such as graphene and transition metal dichalcogenides. However, the valley coherence time, a key quantity for manipulating the valley pseudospin, has never been measured in any material. In this work, we use a sequence of laser pulses to resonantly generate a coherent superposition of excitons (Coulomb-bound electron-hole pairs) in opposite valleys of monolayer WSe$_2$. The imposed valley coherence persists for approximately one hundred femtoseconds. We propose that the electron-hole exchange interaction provides an important decoherence mechanism in addition to exciton population decay. Our work provides critical insight into the requirements and strategies for optical manipulation of the valley pseudospin for future valleytronics applications.**

Group-VI transition metal dichalcogenides (TMDs) with 2H structure (*e.g. MX$_2$, M* = Mo, W; *X* = S, Se) are a particularly intriguing class of semiconductors when thinned down to monolayers.[1,2] The valence and conduction band extrema are located at both *K* and *K*′ points at the corners of the hexagonal Brillouin zone as illustrated in Fig. 1a. The degenerate *K* and *K*′ points are related to each other by time reversal symmetry and give rise to the valley degree of freedom (DoF) of the band-edge electrons and holes. Strong Coulomb interactions lead to the formation of excitons with remarkably large binding energies due to the heavy effective mass and reduced dielectric screening in monolayer TMDs.[3–5] An exciton as a bound electron-hole pair inherits the valley DoF. Because of valley dependent optical selection rules they can be excited only by $\sigma^+$ ($\sigma^-$) circularly polarized light at the $K$ ($K'$) valley. Due to its close analogy to spin,[6] the valley DoF can be considered as a pseudospin represented by a vector **S** on the Bloch sphere (Fig. 1b). The out-of-plane component $S_z$ and in-plane component $S_{x,y}$ describe the polarization and the coherent superposition of exciton valley states. After optical initialization, valley depolarization and decoherence are illustrated by a reduction in the magnitudes of $S_z$ and $S_{x,y}$, respectively.

The ability to coherently manipulate spins and pseudospins is at the heart of spintronics and valleytronics;[7,8] however, previous investigations have mainly focused on the creation and relaxation of valley polarization using non-resonant photoluminescence (PL) or pump/probe spectroscopy techniques.[9–14] Optical excitation close to the lowest energy exciton resonance leads to nearly 100% valley polarization in monolayer TMDs such as $MoS_2$.[9,14,15] Time-resolved PL spectroscopy has revealed a few-picosecond valley polarization decay time, possibly limited by the temporal resolution of the technique.[12] Experiments based on pump/probe spectroscopy reported similar time-scales; however, these measurements may be difficult to interpret due to the fact that only the incoherent exciton population dynamics are probed, which can be sensitive to scattering between optically bright and dark excitons.[16] Even more intriguing are experiments that seem to show that exciton valley coherence—coherent superposition of excitons in $K$ and $K'$ valleys manifested by linearly polarized luminescence—is preserved in PL following non-resonant linearly polarized optical excitation.[17,18] However, steady-state PL does not reveal the details of the valley coherence dynamics.

Directly measuring the timescale over which quantum coherence in the valley pseudospin DoF is preserved remains an outstanding challenge in the field of valleytronics. Exciton valley coherence is a type of non-radiative quantum coherence, *i.e.* coherence between states that are not dipole coupled. Probing exciton valley coherence therefore requires measurements that go beyond traditional linear spectroscopy techniques. In this paper, we examine exciton valley coherence dynamics in monolayer $WSe_2$ using polarization-resolved optical two-dimensional coherent spectroscopy (2DCS).[19] Using a sequence of laser pulses resonant with the exciton transition, we initialize and probe exciton valley coherence, finding that it decays after ~100 fs. The coherence time is faster than the exciton population recombination lifetime—also occurring on a sub-picosecond time scale—pointing to the presence of additional valley decoherence channels.

Following earlier work, we identify the electron-hole exchange interaction as an important decoherence mechanism unrelated to exciton recombination.[20–22] Calculations taking the exchange interaction and the momentum-space distribution of excitons into account reproduce the measured valley coherence dynamics.

We examine monolayer WSe$_2$ flakes ~20 μm in lateral size grown on a sapphire substrate using chemical vapor deposition (See Supplementary Note 1 and Supplementary Fig. 1).[23] Steady-state PL measurements are first performed to identify the exciton resonance and confirm that a high degree of valley polarization can be achieved using circularly polarized optical excitation tuned to 660 nm. We show in Fig. 2a PL spectra for co-circularly ($\sigma^+/\sigma^+$, blue curve) and cross-circularly ($\sigma^+/\sigma^-$, red curve) polarized excitation and detection. The spectra feature a high-energy peak associated with the $A$-exciton ($X$) and a lower-energy peak that we attribute to defect or impurity bound states ($L$). The laser spectrum for the coherent nonlinear optical experiments presented later is indicated by the dashed curve. Following $\sigma^+$ polarized excitation, the PL emission intensity $I_{\sigma+/\sigma-}$ is primarily $\sigma^+$ polarized corresponding to a degree of exciton valley polarization $P_C = \frac{(I_{\sigma+} - I_{\sigma-})}{(I_{\sigma+} + I_{\sigma-})} = 70 \pm 2$ %. The high degree of circular polarization indicates robust initialization of excitons in either valley. We note that the degree of valley polarization in PL measurements varies significantly depending on the excitation wavelength. Thus, it cannot be used to infer valley polarization in the following experiments in which excitons are created resonantly. This PL measurement mainly serves as a familiar characterization method.

In the excitation picture, excitons at $K$ and $K'$ can be modeled as a three-level V system as shown in Fig. 2b. The exciton population decay rate in the $K$ ($K'$) valley is given by $\Gamma_K$ ($\Gamma_{K'}$) and reflects both radiative and non-radiative recombination. The valley coherence time (inversely proportional to $\gamma_v$) is limited by the lifetime of the exciton in either the $K$ or $K'$ valley and can be further reduced in the presence of additional dephasing mechanisms. Non-radiative quantum coherence has been investigated previously in several different systems, including atomic gases, semiconductor quantum wells, and quantum dots, and it is responsible for lasing without inversion, electromagnetic induced transparency, and other interesting non-linear phenomena.[24–26] If the two states involved are not energetically degenerate, the non-radiative coherence between them can be detected via quantum beats in either time-resolved PL, pump/probe, or four-wave mixing spectroscopy. In these experiments, the coherence time can be extracted by examining the decay of the oscillation amplitude.[27,28]

In the case of degenerate valley excitons in TMDs, oscillatory quantum beats are absent and additional care must be taken to separate non-radiative valley coherence from exciton recombination dynamics. To meet this challenge, we use optical 2DCS, which is akin to three-pulse four-wave mixing (or photon echo) spectroscopy with the addition of interferometric stabilization of the pulse delays (see Supplementary Note 2).[29] 2DCS experiments are performed using a sequence of three phase-stabilized laser pulses separated by delays $t_1$ and $t_2$ (see Figs. 2c and 2d, Methods, and Supplementary Note 2). The coherent interaction of the pulses with the sample generates a four-wave mixing signal field $\mathbf{E}_S(t_1, t_2, t_3)$, which is spectrally resolved through heterodyne detection with a fourth phase-stabilized reference pulse. The population recombination and valley coherence dynamics are measured by recording $\mathbf{E}_S$ as the delay $t_2$ is stepped with interferometric precision while the delay $t_1$ is held fixed. Fourier transformation of the signal with respect to $t_2$ yields a two-dimensional coherent spectrum, $\mathbf{E}_S(t_1, \hbar\omega_2, \hbar\omega_3)$, which correlates the "one-quantum" exciton emission energy during delay $t_3$ with the "zero-quantum" energy of the system during the delay $t_2$. Both population relaxation and non-radiative coherence evolution occur during the time delay $t_2$. The experiments are performed at 10 K and low average excitation power to minimize exciton-phonon and exciton-exciton interaction broadening.[30]

Experiments using pulse sequences with carefully chosen polarization combinations provide information on both exciton population relaxation and valley coherence. We first present a 2D spectrum in which all of the excitation pulses and the detected signal are co-circularly polarized ($\sigma^+$). In this case, the first two pulses create an exciton population in the $K$ valley that decays with a rate given by $\Gamma_K$. After a delay $t_2$, the third pulse transfers the remaining population to a configuration with optical coherence between the exciton and ground states that radiates as the four-wave mixing signal field $\mathbf{E}_S$. The spectrum is shown in Fig. 3a, where the exciton emission energy ($\hbar\omega_3$) and zero-quantum energy ($\hbar\omega_2$) are on the horizontal and vertical axes, respectively. The spectrum features a single peak at an emission energy of 1720 meV, consistent with the exciton resonance in the photoluminescence spectrum in Fig. 2a. The peak is centered at $\hbar\omega_2 = 0$ meV since the system is excited to an incoherent population state in the $K$ valley, i.e. the phase does not evolve during the delay $t_2$. A slice along the zero-quantum energy axis is shown in Fig. 3b (dashed curve), which is fit by a square root Lorentzian function[31]. The half-width at half-maximum of the fit provides a measure of $\Gamma_K$, which is equal to 3.4±0.2 meV ($T_1 = \hbar/\Gamma_K = 190\pm10$ fs), consistent with previous lifetime measurements of CVD-grown WSe$_2$. Both radiative and non-radiative recombination contribute to the population decay.[30]

We next probe the exciton valley coherence using a polarization scheme in which the first and third pulses are co-circularly polarized ($\sigma^+$) and the second pulse and detected signal are co-circularly polarized but with opposite helicity ($\sigma^-$). As illustrated in Fig. 2d, the first pulse creates a coherent superposition between the ground and exciton states in the $K$ valley. After a time $t_1$, the second pulse transfers this optical coherence to a coherent superposition between excitons in the $K$ and $K'$ valleys, the evolution of which is monitored during the time $t_2$. The third pulse converts the non-radiative valley coherence to an optical coherence of the $K'$ valley exciton, which is then then detected. The corresponding 2D spectrum is shown in Fig. 3c. The spectrum features a single peak similar to the co-circular excitation sequence; however, the linewidth along $\hbar\omega_2$ is significantly wider as illustrated by the lineshape shown in Fig. 3d. Since the system is in a coherent superposition of valley exciton states during $t_2$, the HWHM linewidth along $\hbar\omega_2$ corresponds to the valley coherence decay rate $\gamma_v$. We measure $\gamma_v = 6.9\pm0.2$ meV ($\tau_v = \hbar/\gamma_v = 98\pm5$ fs)—a factor of two faster than the population recombination rate.

To gain an in-depth understanding of possible processes responsible for exciton valley decoherence, we consider the Maialle-Silva-Sham mechanism—valley relaxation induced by the electron-hole exchange interaction[20,21]. In monolayer TMDs, the exciton valley DoF is intrinsically coupled to the exciton center-of-mass motion by the exchange interaction. The exciton states in opposite valleys are degenerate at zero center-of-mass momentum, but this two-fold valley degeneracy is lifted at finite momentum by the inter-valley exchange interaction, which acts as an in-plane effective magnetic field $\mathbf{\Omega}$ coupled to the valley DoF (Fig. 4a). The magnitude of $\mathbf{\Omega}$ scales linearly with the magnitude of the momentum, while the orientation of $\mathbf{\Omega}$ rotates $4\pi$ when the momentum encloses its origin once. The interplay of intra- and inter-valley exchange interactions leads to unusual exciton dispersion.[32] As shown in Fig. 4a, the lower (upper) exciton band has quadratic (linear) dispersion with the exciton valley pseudospin antiparallel (parallel) to $\mathbf{\Omega}$.

The time dynamics of the valley pseudospin vector $\mathbf{S}$ is described by:

$$\frac{d\mathbf{S}(\mathbf{Q},t)}{dt} = \mathbf{\Omega}(\mathbf{Q}) \times \mathbf{S}(\mathbf{Q},t) + \sum_{\mathbf{Q}'} W_{\mathbf{Q}\mathbf{Q}'}[\mathbf{S}(\mathbf{Q}',t) - \mathbf{S}(\mathbf{Q},t)] - \frac{\mathbf{S}(\mathbf{Q},t)}{\tau}, \qquad (1)$$

where **Q** denotes the center-of-mass momentum. The first term on the right hand side describes precession of **S** around the effective magnetic field **Ω**. The second term captures momentum scattering by a smooth impurity potential which changes the momentum of an exciton but does not act on spin or valley flavor. The last term in equation (1) phenomenologically accounts for scattering events that reduce the information carried by the valley DoF; $\hbar/\tau \equiv \Gamma_K + 2\gamma_K^*$, where $\gamma_K^*$ is the pure dephasing rate of optical coherence between the *K*-valley exciton and the ground state. We assume that **S** initially points along the *x*-direction corresponding to coherence between exciton valley states generated by the first two circularly polarized pulses in the 2DCS experiments.

The equation of motion for **S** is solved numerically. In Fig. 4b the time evolution of the exciton population *N*, with the population decay rate $\Gamma_K$ taken from the co-circular 2DCS experiment, is plotted as the dashed line. The in-plane component $S_x$ averaged over momenta is illustrated by the solid lines in Fig. 4b, which represent coherence between the two valley exciton states. It is clear that the Maialle-Silva-Sham mechanism captures the experimental observations: the valley coherence decays faster than the population relaxes. This result can be understood by recognizing that the effective magnetic field **Ω(Q)** has strong momentum dependence and thus momenta averaging leads to decoherence. The dependence of exciton valley coherence on the momentum scattering rate $\hbar/\tau_p$, which is associated with the scattering strength *W* in equation (1), is also illustrated in Fig. 4b. Surprisingly, our calculations indicate that the larger $\hbar/\tau_p$, the longer the valley coherence time. This counterintuitive behavior is analogous to the motional narrowing effect seen in other solid-state and molecular systems.[33,34] When an exciton changes its momentum more frequently by impurity scattering, the time-averaged effective magnetic field it experiences is reduced. Quantitative agreement between the calculations and experiment is obtained for a momentum scattering rate $\hbar/\tau_p \cong 10$ meV. We note that impurity scattering enhances both the momentum scattering rate $(\hbar/\tau_p)$ and the exciton pure dephasing rate $(\gamma_K^*)$, the latter of which suppresses the valley coherence time. Interestingly, additional 2DCS measurements at elevated temperatures reveal that the valley coherence decay rate $\gamma_v$ increases to 7.5±0.2 meV at 30 K, whereas the population recombination rate $\Gamma_K$ remains unchanged within the estimated uncertainty (data not shown). This weak temperature dependence suggests that acoustic phonon scattering plays a small role in exciton valley decoherence and is additional evidence that the electron-hole exchange interaction provides an efficient valley decoherence mechanism.

The experimental and theoretical results presented here offer critical insight into valley decoherence mechanisms. Such information also provides important guidance for implementing valleytronics based on TMDs because all coherent manipulations of the exciton valley pseudospin have to be performed before coherence is lost. Several important messages regarding valley physics in monolayer TMD emerge from this work. First, understanding the mechanisms limiting the exciton population lifetime is of utmost importance since exciton recombination places an upper bound on the valley coherence time. Second, impurity scattering introduces competing effects on exciton valley coherence. While certain types of impurities may lead to non-radiative recombination and shorten the population lifetime, impurity scattering can also *enhance* coherence by effectively screening the exchange interaction. Finally, in our experiments we resonantly excite excitons confined within a single monolayer, which are expected to exhibit intrinsic rapid radiative decay. Extending the exciton valley coherence time might be possible through resonant optical excitation of alternative exciton-type states. For example, strongly localized quantum-dot-like exciton states[35,36] in monolayers and indirect excitons in multi-layer heterostructures[37] couple more weakly with the optical field and thus exhibit a longer population lifetime. Exciting these states may allow for coherent control of the valley pseudospin DoF over a longer duration. Alternatively, embedding TMDs in photonic cavities creates hybrid exciton-polariton states with lifetimes that can be tuned by controlling the exciton-photon coupling parameter[38]. These advanced materials and photonic structures will likely offer additional flexibility in controlling the valley DoF.

## Methods

**Sample Preparation and Characterization**

The monolayer $WSe_2$ sample was grown using chemical vapor deposition as described in detail in Ref. (23) and Supplementary Note 1. Briefly, the sample was synthesized on a double-side polished sapphire substrate for transmission optical spectroscopy experiments. We verified the monolayer thickness using atomic force microscopy. The sample is mounted in a liquid helium cold-finger cryostat and held at a temperature of 10 K. The photoluminescence experiments were performed using a continuous wave laser at 660 nm, which was focused to a spot size of ~2 μm in diameter. The photoluminescence spectra in Fig. 2a were fit with double Gaussian functions to determine the degree of circular polarization.

**Two-Dimensional Coherent Spectroscopy**

100-fs pulses generated from a mode-locked Ti:sapphire laser at a repetition rate of 80 MHz are split into a set of four phase-stabilized pulses. Three of the pulses are focused to an overlapped 35 µm FWHM spot on the sample, which likely probes a few monolayer flakes. The coherent interaction of the three fields with the sample generates a third-order nonlinear optical signal field, $\mathbf{E}_S(t_1, t_2, t_3)$, which is a photon echo (or four-wave mixing signal) detected in transmission in the wavevector-matching direction. $\mathbf{E}_S$ is interferometrically measured via heterodyne detection with a fourth phase-stabilized reference field $\mathbf{E}_R$. Both phase and amplitude of $\mathbf{E}_S$ as a function of emission frequency $\hbar\omega_3$ are extracted from the spectral interferograms of $\mathbf{E}_S$ and $\mathbf{E}_R$ obtained using a spectrometer. Subsequent Fourier transformation of the signal field with respect to the delay $t_2$ yields a rephasing "zero-quantum" spectrum with amplitude given by $\mathbf{E}_S(t_1, \hbar\omega_2, \hbar\omega_3)$. We use a value of $t_1 = 0$ fs to obtain maximum signal-to-noise; however using a value of $t_1 = 100$ fs, which is larger than the pulse duration, provides similar results other than an overall smaller signal amplitude. The pump fluence at the sample is kept below 1 µJ cm$^{-2}$, which corresponds to an excitation density of ~2 × 10$^{11}$ excitons cm$^{-2}$.

**Acknowledgements** The theoretical and experimental collaboration is made possible via SHINES, an Energy Frontier Research Center funded by the U.S. Department of Energy (DoE), Office of Science, Basic Energy Science (BES) under award # DE-SC0012670. K.H., F.W., L.X., X.L., and A.H.M. have all received support from SHINES. Optical spectroscopy studies performed by K.H., C.K.D., and X.L. have been partially supported by NSF DMR-1306878 and Welch Foundation F-1662. A.H.M. also acknowledges support from Welch Foundation F-1473. L.J.L. thanks support from KAUST Saudi Arabia, Academia Sinica Taiwan, and AOARD-134137 USA. C.H.C. thanks support from Ministry of Science and Technology Taiwan (MOST 104-2218-E-035-010 and 104-2628-E-035-002-MY3).

**Author Contributions** K.H and G.M. contributed equally to this work. G.M. and X.L. conceived the concept. K.H. led the experimental effort. All co-authors at the University of Texas ran the experiments, acquired the data, and analyzed the results. C.-H.C., M.-Y.L., and L.-J.L. provided the samples. F.W. and A.H.M. performed the theoretical studies. G.M., F.W., and X.L. wrote the manuscript. All authors discussed the results and commented on the manuscript at all stages.

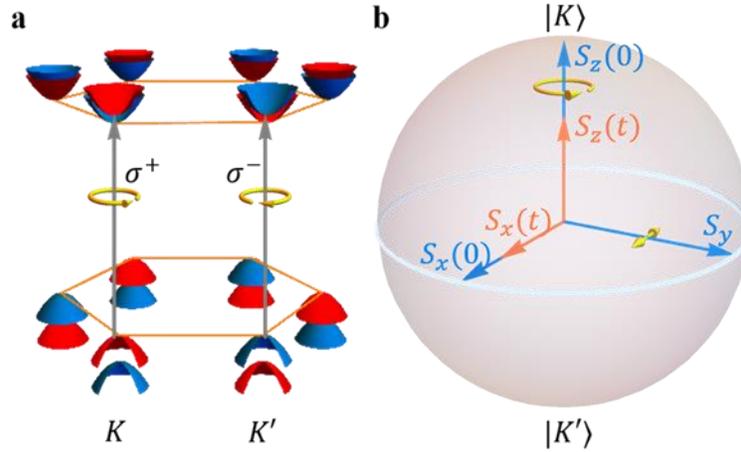

**Figure 1: Coupled spin and valley degrees of freedom at the band extrema.** (a) Schematic band structure and optical selection rules in monolayer TMDs. Left-circularly polarized ($\sigma^+$) and right-circularly polarized ($\sigma^-$) light couples the valence and conduction bands at the $K$ and $K'$ valleys, respectively. (b) Bloch sphere representation of the valley pseudospin DoF. Exciton valley polarization and coherence are described by vectors $S_z$ and $S_{x,y}$, respectively. Following initial optical excitation, valley depolarization and decoherence result in a decrease in the magnitude of $S_z$ and $S_{x,y}$ with time, respectively.

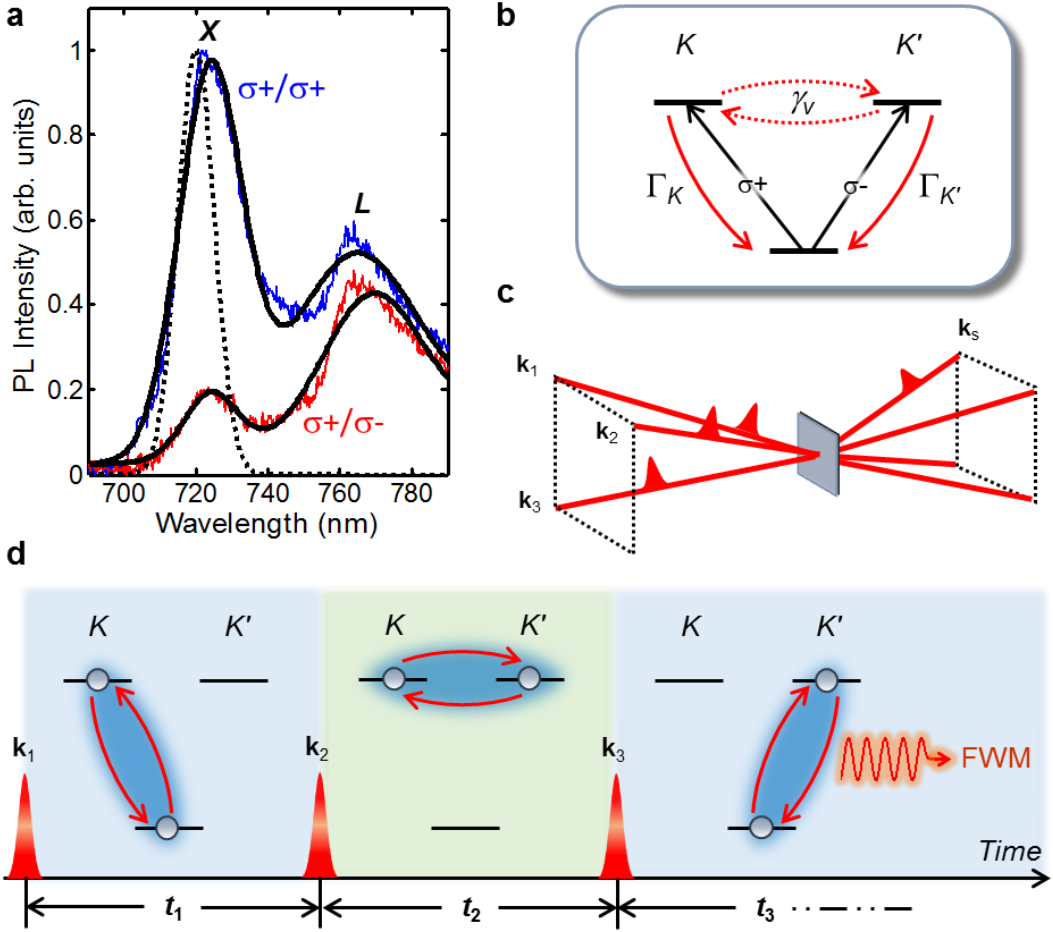

**Figure 2: Ultrafast resonant generation and detection of exciton valley coherence in monolayer TMDs.** (**a**) Low temperature PL spectra of monolayer WSe$_2$ using σ+ polarized, 660 nm excitation for co- (σ+) and cross-circularly (σ-) polarized detection. The spectra feature two peaks associated with the *A*-exciton (*X*) and defect or impurity bound states (*L*). The laser spectrum used for the nonlinear experiments is indicated by the dashed line. (**b**) The crystal ground state and two exciton states in the *K* and *K'* valleys can be modeled by the three-level V system in the excitation picture. An exciton population in the *K* (*K'*) valley recombines with a rate $\Gamma_K$ ($\Gamma_{K'}$). Coherence between the two exciton transitions decays with the valley coherence rate given by $\gamma_v$. (**c**) 2DCS experiment schematics. Three phase-stabilized pulses are focused onto the monolayer WSe$_2$ sample with wavevectors **k**$_1$, **k**$_2$, and **k**$_3$. The nonlinear interaction generates a four-wave mixing (FWM) signal that is radiated in the wave-vector phase-matched direction **k**$_s$. (**d**) Illustration of the resonant generation and detection of exciton valley coherence. The first laser pulse creates a coherent superposition between the *K*-valley exciton and ground states (left panel), which is transferred to a non-radiative exciton valley coherence by the second pulse (middle panel). The third pulse converts the valley coherence to an exciton coherence in the *K'* valley (right panel), which is optically detected.

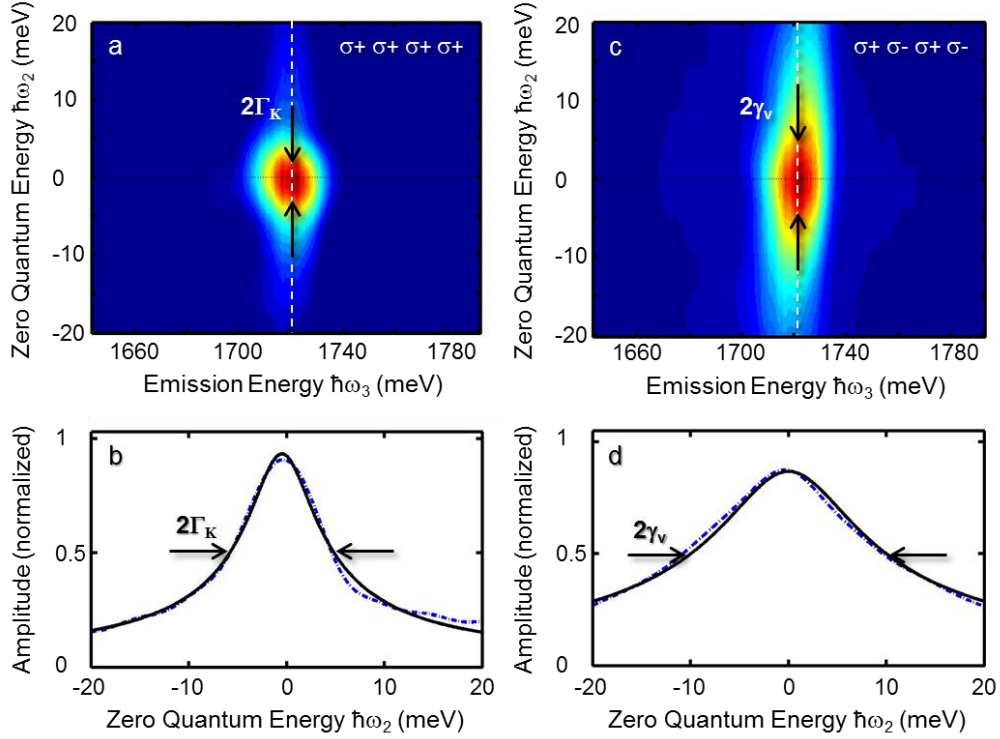

**Figure 3: Exciton population relaxation and valley coherence measured with 2DCS.** **(a)** The exciton population relaxation rate $\Gamma_K$ can be extracted from the 2D spectrum using co-circular polarization ($\sigma^+$) of all excitation pulses and detected signal. The amplitude spectrum features a single peak at $\hbar\omega_3 = 1720$ meV associated with the emission of excitons in the $K$ valley. The half-width at half-maximum along the zero-quantum energy axis $\hbar\omega_2$ provides a measure of the exciton population decay rate $\Gamma_K$. A slice along the white dashed line is shown in **(b)**. **(c)** When using alternating helicity of the excitation pulses and detected signal, the system is driven into a coherent superposition between the $K$ and $K'$ valley excitons. The width of the lineshape along $\hbar\omega_2$, shown by the slice in **(d)**, provides a measure of the valley quantum coherence decay rate $\gamma_v$.

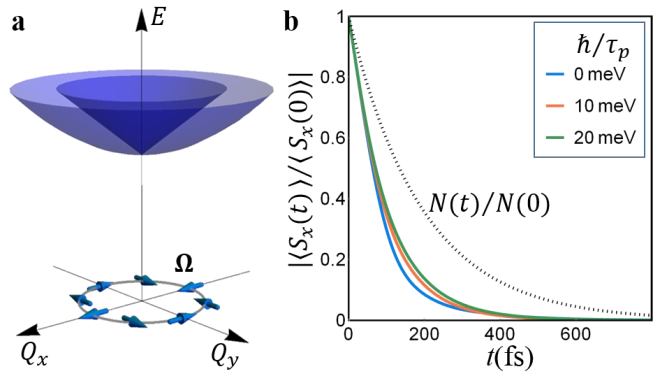

**Figure 4: Electron-hole exchange coupling between valley excitons.** (**a**) Exciton valley bands in momentum space are illustrated. The electron-hole inter-valley exchange interaction acts as an effective magnetic field $\mathbf{\Omega}$ that is responsible for relaxation of exciton valley coherence. The orientation of $\mathbf{\Omega}$ depends on the exciton momentum $Q_{x,y}$. (**b**) When averaging over excitons distributed in momentum space, valley decoherence (solid lines) is faster than the population recombination lifetime $N(t)/N(0)$ (dashed line). A larger exciton momentum scattering rate ($\tau_p$) results in slower valley decoherence.

# Supplementary Information for: Direct Measurement of Exciton Valley Coherence in Monolayer WSe$_2$

Kai Hao[1,*], Galan Moody[1,†,*], Fengcheng Wu[1], Chandriker Kavir Dass[1], Lixiang Xu[1], Chang-Hsiao Chen[2], Ming-Yang Li[3], Lain-Jong Li[3], Allan H. MacDonald[1] and Xiaoqin Li[1]

[1] Department of Physics and Center for Complex Quantum Systems, University of Texas at Austin, Austin, TX 78712, USA.

[2] Department of Automatic Control Engineering, Feng Chia University, Taichung 40724, Taiwan.

[3] Physical Science and Engineering Division, King Abdullah University of Science & Technology (KAUST), Thuwal 23955, Saudi Arabia.

[†]Present address: National Institute of Standards & Technology, Boulder, CO 80305, USA.

[*]These authors contributed equally to this work.

## Supplementary Note 1

*Monolayer WSe$_2$ Sample*: CVD monolayer WSe$_2$ triangular crystals were synthesized based on previous work[1]. In brief, a double-side polished sapphire (0001) substrate (from *Tera Xtal Technology Corp.*) was cleaned in a H$_2$SO$_4$/H$_2$O$_2$ (70:30) solution heated at 100 °C for one hour. After cleaning, the sapphire substrate was placed on a quartz holder in the center of a 1″ tubular furnace. Precursor of 0.3 grams WO$_3$ powder was placed in the heating zone center of the furnace (99.5% from *Sigma-Aldrich*). Se powder (99.5% from *Sigma-Aldrich*) was placed in a quartz tube at the upstream position of the furnace tube, which was maintained at 270 °C during the reaction. The sapphire substrate was located at the downstream side, where the WO$_3$ and Se vapors were brought into contact with the sapphire substrate by an Ar/H$_2$ flowing gas (H$_2$ = 20 sccm, Ar = 80 sccm, chamber pressure = 3.5 Torr). The reaction heating zone was held at 925 °C (ramping rate of 25 °C/min). The actual temperature of the sapphire substrate was ranged from 750 °C to 850 °C. The heating zone was held at 925 °C for 15 minutes after which the furnace was then naturally cooled to room temperature. The reaction yielded monolayer WSe$_2$ flakes triangular in shape with a base width of ~20 μm. The thickness was determined using atomic force microscopy, shown in Supplementary Fig. 1, which confirms the monolayer thickness of the sample.

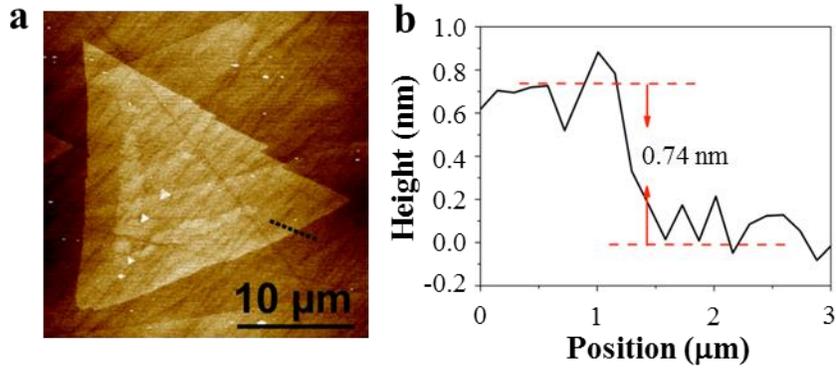

**Supplementary Figure 1: Atomic force microscope image of monolayer WSe$_2$. (a)** Individual triangular monolayer flakes are visible in the atomic force microscope image. A slice along the dashed line is shown in **(b)**, illustrating the monolayer thickness of the flakes.

*Linearly Polarized Photoluminescence:* Linearly polarized photoluminescence (PL) from monolayer WSe$_2$ is measured using a 660 nm laser excitation source. Supplementary Fig. 2 shows polar plots of the exciton PL peak intensity ($r$) as a function of the detected angle ($\vartheta$) for two excitation polarization direction (indicated by the arrows). Linearly polarized PL following the polarization of the excitation laser has been used as the key experimental evidence for valley coherence in previous studies[2]. We repeat this linearly polarized PL experiment to demonstrate the consistent properties for the monolayer WSe$_2$ investigated in our nonlinear experiments. The solid lines are fits using the equation $r = A \times (1 + B \times sin(\vartheta - \varphi))$, where $\varphi$ is the polarization angle corresponding to maximum exciton PL intensity. The results in Supplementary Fig. 2 demonstrate that $\varphi$ is determined entirely by the excitation laser polarization angle and is independent of crystal orientation. For both excitation angles, we measure a degree of linear polarization $P_L = \frac{(I_H - I_V)}{(I_H + I_V)} = \approx 30\%$, where $I_H$ ($I_V$) is the PL intensity parallel (perpendicular) to the excitation laser polarization direction. PL spectra were collected in the range of 0 to 180 degree in the detection angle, and spectra are duplicated in the range of 180 to 360 degree in the polar plot.

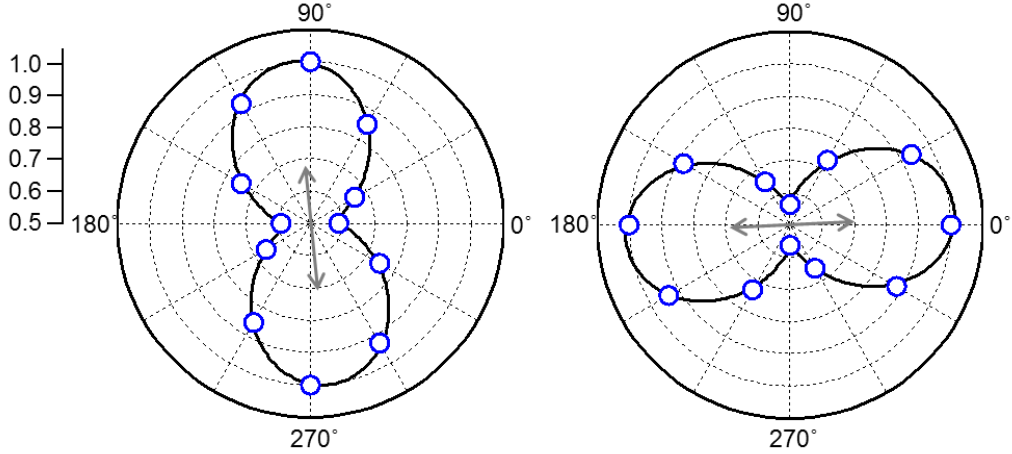

**Supplementary Figure 2: Linearly polarized photoluminescence from monolayer WSe$_2$.** Normalized photoluminescence peak intensity ($r$) as a function of detection angle ($\vartheta$) for a given excitation laser polarization (indicated by the arrows).

*Two-Dimensional Coherent Spectroscopy*: Optical two-dimensional coherent spectroscopy (2DCS) is an enhanced version of three-pulse four-wave mixing in which the pulse delays are varied with interferometric precision. A sequence of three pulses with variable delays is generated using a set of folded and nested Michelson interferometers. The pulses are generated by a mode-locked Ti:sapphire laser operating at an 80 MHz repetition rate with a pulse duration of ~100 femtoseconds. This setup enables femtosecond control of the pulse delays with a stabilization of $\lambda/100$, which allows for the four-wave mixing signal to be Fourier transformed. Analysis of the signal in the Fourier spectral domain, combined with phase cycling of the pulse delays to minimize scatter of the laser pulses into the spectrometer, suppresses optical frequency noise overlapping with the exciton resonance and enables us to isolate the population recombination and valley coherence signals.

In the experiments, three of the pulses with wavevectors $\mathbf{k}_1$, $\mathbf{k}_2$, and $\mathbf{k}_3$ are focused to a single 35 μm spot on the sample that is held at a temperature of 10 K in an optical cryostat. The interaction of the first pulse with the sample with wavevector $\mathbf{k}_1$ excites an electronic coherence between the exciton ground and excited states. The second pulse with wavevector $\mathbf{k}_2$ converts the optical coherence into a transient population grating (in the case of co-circular polarization) or a valley coherence grating (in the case of alternating helicity of the pulses). After a time $t_2$, the third pulse with wavevector $\mathbf{k}_3$ generates an optical coherence that diffracts off the grating and is detected in transmission in the wavevector-matching direction $\mathbf{k}_s = -\mathbf{k}_1+\mathbf{k}_2+\mathbf{k}_3$. This four-wave mixing signal is interferometrically and spectrally resolved using a fourth phase-stabilized reference pulse while the delay $t_2$ is varied. Subsequent Fourier transformation of the signal yields a rephasing zero-quantum spectrum with amplitude given by $\mathbf{E_S}(t_1, \hbar\omega_2, \hbar\omega_3)$. In the present experiments, we set $t_1 = 0$ fs to obtain maximum signal-to-noise; however using a value of $t_1 = 100$ fs, which is equal to the pulse duration, does not lead to qualitative difference in the data other than an overall smaller amplitude.

Polarization control of the excitation pulses and detected signal allow for isolation of valley coherence dynamics from population recombination. The coherent light-matter interaction can be modeled by the density matrix, $\rho$. Here, the diagonal terms of $\rho$ represents the exciton population in valley $K$ and $K'$, while the off-diagonal terms describe exciton coherence. Using perturbation theory in the applied field[3], the three-pulse excitation scheme of the $K$-valley exciton for co-circular polarization is described by the following sequence:

$$\rho_{00} \xrightarrow{\mathbf{E}_1(0)} \rho_{0K} \xrightarrow{\mathbf{E}_2(t_1)} \rho_{KK} \xrightarrow{\mathbf{E}_3(t_2)} \rho_{K0} \xrightarrow{\mathbf{E}_S(t_3)} \rho_{00}, \qquad (1)$$

where $\mathbf{E}_i(t)$ corresponds to the $i^{th}$ excitation pulse and $t$ is the pulse arrival time. It is clear from supplementary equation (1) that the first two pulses excite an exciton population in the $K$ valley; therefore by recording the four-wave mixing signal while scanning the delay $t_2$, we can extract the exciton population recombination rate $\Gamma_K$.

To resonantly excite and detect the exciton valley coherence, we use a pulse excitation scheme in which the first and third pulses are co-circularly polarized ($\sigma^+$) and the second pulse and detected signal are co-circularly polarized but with opposite helicity ($\sigma^-$). This scheme can be described by the following density matrix sequence:

$$\rho_{00} \xrightarrow{\mathbf{E}_1(0)} \rho_{0K} \xrightarrow{\mathbf{E}_2(t_1)} \rho_{K'K} \xrightarrow{\mathbf{E}_3(t_2)} \rho_{K'0} \xrightarrow{\mathbf{E}_S(t_3)} \rho_{00}. \qquad (2)$$

As shown in Fig. 2d of the main text, the first pulse generates a coherent superposition between the ground and exciton state in the $K$ valley. After a time $t_1$, the second pulse transfers this optical coherence to a non-radiative coherence between excitons in the $K$ and $K'$ valleys, the evolution of which we monitor during the time $t_2$. The third pulse converts the valley coherence to an optical coherence in the $K'$ valley, which is detected as the radiated four-wave mixing signal field. Thus by measuring the four-wave mixing signal while scanning the delay $t_2$, we directly probe the valley coherent dynamics and dephasing rate $\gamma_v$.

**Supplementary Note 2**

*Theory of exciton valley dynamics*: We start from the exciton problem in monolayer TMDs. The band extrema in valley $K$ or $K'$ can be described by the massive Dirac model, given by

$$\mathcal{H}_k = \begin{pmatrix} \Delta & \hbar v_F(\pm k_x + i\, k_y) \\ \hbar v_F(\pm k_x - i\, k_y) & -\Delta \end{pmatrix}, \tag{3}$$

where the sign $+/-$ corresponds to valley $K/K'$. To account for the finite thickness of monolayer TMDs, we use the following form of electron-hole interaction potential[4]:

$$V_q = \frac{2\pi e^2}{\epsilon q}\frac{1}{1+r_0 q}, \tag{4}$$

where $\epsilon$ is the environment-dependent dielectric constant and $r_0$ is a parameter that is inversely proportional to $\epsilon$.

The characteristic length and energy scale are respectively the effective Bohr radius and Rydberg energy, given by

$$a_B^* = \frac{2\epsilon(\hbar v_F)^2}{e^2 \Delta}, \quad \mathrm{Ry}^* = \frac{e^2}{2\epsilon a_B^*}. \tag{5}$$

The effective fine structure can be defined as $\alpha = e^2/(\epsilon \hbar v_F)$.

For monolayer WSe$_2$, $\hbar v_F = 3.310\text{Å} \times 1.19\text{eV}$ and $\Delta = 0.685\text{eV}$, which are taken from Ref. *(5)*. We note that this value of $\Delta$ obtained from DFT calculations underestimates the band gap, which, however, should not noticeablely affect the quantities we are interested in for this study. We adjust the parameter $r_0$ so that the *A*-exciton binding energy for $\epsilon = 2.5$, which corresponds to WSe$_2$ lying on an SiO$_2$ substrate and exposed to air, is 0.370 eV as measured in the recent experiments.[6] We therefore find $r_0 = 22.02\text{Å}/\epsilon$ by solving the Bethe-Salpeter equation for excitons in the massive Dirac model.[7]

In monolayer TMDs, the center-of-mass motion of an exciton is coupled to its valley degree of freedom as described by the Hamiltonian

$$H_Q = (\hbar\omega_0 + T_Q + J_Q)\sigma_0 + J_Q[\cos(2\phi_Q)\sigma_x + \sin(2\phi_Q)\sigma_y], \qquad (6)$$

where $\sigma_0$ and $\sigma_{x,y}$ are identity matrix and Pauli matrices in the exciton valley space, respectively, $\hbar\omega_0$ is the excitation energy of the A exciton, $T_Q = \hbar^2 Q^2/2M$ is the kinetic energy of the center-of-mass motion with $M$ being the total mass of the exciton, and $\phi_Q$ is the orientation angle of the center-of-mass momentum $\mathbf{Q}$. The intra and inter-valley exchange interaction are described by the $J_Q\sigma_0$ and $\sigma_{x,y}$ terms, respectively. The coupling constant $J_Q$ is linear in the magnitude of $\mathbf{Q}$:

$$J_Q = \text{Ry}^* \frac{\pi}{4}\alpha^2 |a_B^*\psi(0)|^2 (2T_Q/\text{Ry}^*)^{1/2}, \qquad (7)$$

where $|a_B^*\psi(0)|^2$ is the probability that an electron and a hole spatially overlap.[8]

For WSe$_2$ lying on sapphire substrate and exposed to air, the dielectric constant $\epsilon$ is approximately 5.5. The appropriate parameter values are then $\text{Ry}^* = 75.66$ meV, $\alpha = 0.66$, and $|a_B^*\psi(0)| = 0.88$. Therefore, $J_Q = 20 \text{ meV} \times (2T_Q/\text{Ry}^*)^{1/2}$.

The time evolution of excitons is governed by:

$$\frac{d\rho(\mathbf{Q},t)}{dt} = \frac{i}{\hbar}[\rho(\mathbf{Q},t), H_Q] + \sum_{Q'} W_{QQ'}[\rho(\mathbf{Q'},t) - \rho(\mathbf{Q},t)] - \frac{\rho(\mathbf{Q},t)}{\tau}, \qquad (8)$$

where $\rho(\mathbf{Q},t)$ represents a $2 \times 2$ density matrix in the exciton valley space at momentum $\mathbf{Q}$ and time $t$. The diagonal terms of $\rho$ stand for the exciton population in valley $K$ and $K'$, while the off-diagonal terms describe the coherence between $K$ and $K'$ exciton states. The final term in supplementary equation (8) phenomenologically captures the effects of exciton recombination ($\Gamma_K$) and pure dephasing ($\gamma_K^*$) on valley coherence, where $\hbar/\tau \equiv \Gamma_K + 2\gamma_K^*$. The underlying assumption for the $2\gamma_K^*$ contribution is that the pure dephasing processes for $K$ and $K'$ excitons are uncorrelated.

We decompose the density matrix into $\rho(\mathbf{Q},t) = \frac{1}{2}N(\mathbf{Q},t)\sigma_0 + \mathbf{S}(\mathbf{Q},t)\cdot\boldsymbol{\sigma}$. The equation of motion for the valley pseudospin vector $\mathbf{S}$ makes the physical picture more revealing:

$$\frac{d\mathbf{S}(\mathbf{Q},t)}{dt} = \boldsymbol{\Omega}(\mathbf{Q}) \times \mathbf{S}(\mathbf{Q},t) + \sum_{Q'} W_{QQ'}[\mathbf{S}(\mathbf{Q'},t) - \mathbf{S}(\mathbf{Q},t)] - \frac{\mathbf{S}(\mathbf{Q},t)}{\tau}, \qquad (9)$$

where $\boldsymbol{\Omega}(\mathbf{Q}) = 2J_Q(\cos(2\phi_Q), \sin(2\phi_Q), 0)/\hbar$. We make the assumption of elastic momentum scattering, *i.e.* $W_{QQ'}$ is nonzero only if $|\mathbf{Q}| = |\mathbf{Q'}|$. Thus we can write

$$\frac{d\mathbf{S}(\mathbf{Q},\phi,t)}{dt} = \mathbf{\Omega}(\mathbf{Q},\phi) \times \mathbf{S}(\mathbf{Q},\phi,t) +$$
$$\frac{1}{2\pi}\int_0^{2\pi} d\phi' W(\phi-\phi')[\mathbf{S}(\mathbf{Q},\phi',t) - \mathbf{S}(\mathbf{Q},\phi,t)] - \frac{\mathbf{S}(\mathbf{Q},\phi,t)}{\tau}. \quad (10)$$

To take advantage of the rotational symmetry, we make the angular Fourier transformation:

$$\mathbf{S}(\mathbf{Q},\phi,t) = \sum_n \mathbf{S}^{(n)}(\mathbf{Q},t)e^{in\phi}, \quad (11)$$

$$\hbar \frac{d}{dt}\begin{pmatrix} S_+^{(0)}(\mathbf{Q},t) \\ S_z^{(-2)}(\mathbf{Q},t) \\ S_-^{(-4)}(\mathbf{Q},t) \end{pmatrix} = \begin{pmatrix} -\hbar/\tau & -i\,2J_Q & 0 \\ -iJ_Q & -\hbar/\tau_2^* & iJ_Q \\ 0 & i\,2J_Q & -\hbar/\tau_4^* \end{pmatrix} \begin{pmatrix} S_+^{(0)}(\mathbf{Q},t) \\ S_z^{(-2)}(\mathbf{Q},t) \\ S_-^{(-4)}(\mathbf{Q},t) \end{pmatrix}, \quad (12)$$

where $S_\pm^{(n)} = S_x^{(n)} \pm iS_y^{(n)}$, and $1/\tau_n^* = 1/\tau_n + 1/\tau$ with $1/\tau_n$ being the momentum scattering rate:

$$\frac{1}{\tau_n} = \frac{1}{2\pi}\int_0^{2\pi} d\phi\, W(\phi)[1 - \cos(n\phi)]. \quad (13)$$

We assume δ-function impurity potentials. Then $W(\phi)$ has no $\phi$ dependence. Therefore, $1/\tau_2 = 1/\tau_4 \equiv 1/\tau_p$.

When **S** initially points to the *x*-direction, then $S_x(\mathbf{Q}, t)$ averaged over momenta has the form:

$$\langle S_x(t)\rangle = \frac{1}{2\pi}\int d^2\mathbf{Q}\, S_+(\mathbf{Q},t) = \int d\mathbf{Q}[\mathbf{Q}S_+^{(0)}(\mathbf{Q},t)]. \quad (14)$$

The initial momentum dependence of $S_x$ is modeled by a Lorentzian distribution given by $S_x(\mathbf{Q}, 0) = S_x(0,0)/(1 + 4T_Q^2/\delta^2)$, where $T_Q$ is the kinetic energy associated with the exciton center-of-mass motion. We approximate the parameter $\delta$ by the homogeneous linewidth, assuming that finite-momentum states are able to radiate light only if their energy is within the homogeneous linewidth. This approximation also assumes that excitons outside the light-cone can radiate due to interaction with phonons and impurities. The time evolution of $\langle S_x(t)\rangle$ is shown in Fig. 4 of the main text.